\documentclass[aps,prl,twocolumn,groupedaddress,amsmath,amssymb,amsfonts,graphics]{revtex4}

\usepackage{graphicx}
\usepackage{bm}
\usepackage{bbold}
\usepackage{wasysym}
\usepackage{ulem}
\usepackage{color}
\usepackage{subfigure}

\def\l{\left}
\def\r{\right}
\def\<{\begin{equation}}
\def\>{\end{equation}}
\newcommand{\bra}[1]{\left\langle #1\right|}              

\newcommand{\ket}[1]{\left| #1\right\rangle}

\begin{document}

\title{Kondo Vortices, Zero Modes, and Magnetic Ordering in a Kondo Lattice Model}

\author{Saeed Saremi}
\affiliation{Department of Physics, Massachusetts Institute of Technology, 77 Massachusetts Avenue, Cambridge, MA 02139}
\author{Patrick A. Lee}
\affiliation{Department of Physics, Massachusetts Institute of Technology, 77 Massachusetts Avenue, Cambridge, MA 02139}
\author{T. Senthil	}
\affiliation{Department of Physics, Massachusetts Institute of Technology, 77 Massachusetts Avenue, Cambridge, MA 02139}

\date{March 25, 2009}

\begin{abstract}
Motivated by the mysteries of the heavy fermion quantum critical point, we investigate  the competition between Kondo screening and magnetic ordering in the honeycomb Kondo lattice at half filling. We  examine the destruction of the Kondo phase by proliferating vortex configurations in the Kondo hybridization order parameter. We find that there are zero modes associated with Kondo vortices. Condensing these vortices can lead to the antiferromagnetic phase.
\end{abstract}

\maketitle
The quantum critical point (QCP) separating the anti-ferromagnetic (AF)  phase and the heavy Fermi liquid (HFL) phase which is tuned by either magnetic field, pressure, or chemical doping is a topic of great current interests~\cite{Coleman05, Lohneysen07}. The HFL phase is characterized by its large Fermi surface, in contrast to the AF phase with a small Fermi surface. One suggestion~\cite{Senthil05, Lohneysen07} is that the Kondo screening and magnetic ordering {\it both} collapse at the critical point. Clearly this {\it simultaneous} collapse is the  key physics that has to be understood, and it may lead to understanding the strong non-Fermi liquid behavior observed in the quantum critical region above the QCP. However there are no established theories that explain why this simultaneous collapse should happen and there are no concrete models in which the simultaneous collapse of  Kondo and AF scale is studied.

In this letter we take a step in this direction. Our approach is to identify
configurations in the hybridization mean field theory of the Kondo screening phase which may drive the transition to magnetism while at
the same time kill the hybridization. Natural
candidates are vortex configurations of the complex
mean field hybridization amplitude whose
structure of vortices may be determined within the
hybridization mean field theory. Such vortices are interesting per se as possible
mean field configurations in the  Kondo screened phase and their
core structure might offer clues on the transition to magnetism.
Similar phenomena happen near deconfined quantum critical points in insulating magnets~\cite{Senthiletal}. As a concrete example we study the Kondo-Heisenberg model on a half-filled honeycomb lattice. We show that such a vortex configuration supports zero modes of the fermionic degrees of freedom. This allows the construction of a spin triplet vortex. Proliferation of these vortices suppresses the Kondo phase and simultaneously produces antiferromagnetic order. Despite this apparently exotic route, in this example the critical properties of this transition are in the conventional $O(3)$ universality class.

The Hamiltonian of the Kondo-Heisenberg model on the honeycomb lattice at half-filling is:
\begin{equation} \label{KH-hamiltonian}
\hat{H} = -t \sum_{\langle ij\rangle\alpha} \left( c_{i\alpha}^\dagger c_{j\alpha} + H.c. \right) + J_K \sum_i \bm{s}_i\cdot \bm{S}_i + J_H \sum_{\left<ij\right>} \bm{S}_i\cdot \bm{S}_j,
\end{equation}
where $\alpha \in \{\uparrow, \downarrow\}$ is the spin flavor. We start inside the KI phase. The simplest way to realize the Kondo screening phase is to write the localized spin $\bm{S}_i$ at site $i$ by employing slave fermions $f_i$:
\begin{align}
 \bm{S}_i &= \frac{1}{2} f_{i \alpha}^{\dagger} \bm{\sigma}_{\alpha\beta} f_{i\beta},
\end{align}
where the Hilbert space for the slave fermions is constrained $\sum_\alpha f_{i\alpha}^\dagger f_{i \alpha} =1$. The Kondo screening phase is now realized at the mean field level when the admixture $b_i = \langle f_{i\alpha}^{\dagger}c_{i\alpha} \rangle$ is nonzero and opens a gap in the Dirac spectrum ~\cite{Saremi07}. The Heisenberg term also introduces the following RVB mixing $\chi_{ij} = \langle f_{i\alpha}^{\dagger}f_{j\alpha} \rangle$. Both $b_i=|b_i|e^{i\theta_i}$ and $\chi_{ij}= |\chi_{ij}|e^{ia_{ij}}$ are complex-valued. Beyond mean field theory, $b_i$  is not gauge invariant and never orders but the development of the Kondo amplitude $|\langle b_i \rangle|=b$ provides a stiffness which we are going to further explore here. In particular we consider proliferating Kondo vortex configurations as a way of destroying the Kondo coherence. The Kondo field for a KV in the continuum limit takes the form $b(\bm{r}) = |b(r)| e^{\pm i\theta(\bm{r})}$, where $|b(r)| \propto r$ as $r\rightarrow
 0$ and converges to the mean field value $b_\infty$ as $r\rightarrow \infty$. Due to the presence of $|(\partial_\mu+ ia_\mu) b|^2$ in the action, the finite energy configurations are obtained by inserting a $\mp 2\pi$ gauge flux extended around the vortex core. An example of such a configuration is
\begin{align}
\label{KVB1}b(\bm{r}) &= b_{\infty} {\tanh} (r/\xi_b) e^{+ i\theta(\bm{r})},\\
\label{KVB2}a_\theta (r) &= -{\tanh}^2(r/\Lambda)/r,
\end{align}
where the following gauge $\bm{a}(\bm{r}) = a_\theta (r) \hat{\bm{\theta}}$ is chosen for the gauge field. We stay with this gauge choice for our discussion of the zero mode equations.

The mean field Kondo Hamiltonian is given by
\begin{equation}
\begin{split}
\label{H_2}\hat{H}_{2} &= -t \sum_{\left<ij\right>\alpha}(c_{i\alpha}^\dagger c_{j\alpha}+H.c.)\\&+\chi\sum_{\left<ij\right>\alpha} (e^{i a_{ij}}f_{i\alpha}^\dagger f_{j\alpha}+H.c.)\\&+\sum_{i\alpha}(b_i c_{i\alpha}^\dagger f_{i\alpha} + H.c.),
\end{split}
\end{equation}
where the chemical potentials $\mu^c$ and $\mu^f_i$ for the $c$ and $f$ fermions have been set to zero. It turns out that the average fermion and electron number is maintained at half-filling, even though particle-hole symmetry is destroyed in the presence of vortices.  The proof   proceeds in two steps. I) The simultaneous transformation $\{b_i \rightarrow b_i^*, a_{ij} \rightarrow -a_{ij}\}$ converts a vortex to an anti-vortex. The wavefunctions become complex conjugated and the density, $\langle c_{i\alpha}^\dagger c_{i\alpha} \rangle $ stays the same.  II) Under the additional particle-hole transformation $\{ c_{i\alpha} \rightarrow \epsilon_i c_{i\alpha}^\dagger,~ f_{i\alpha} \rightarrow -\epsilon_i f_{i\alpha}^\dagger,~ b_i \rightarrow b_i^*,~ a_{ij} \rightarrow -a_{ij} \}$,  the Hamiltonian is invariant. This means that $\langle c_{i\alpha}^\dagger c_{i\alpha} \rangle$ for a vortex equals $1-\langle c_{i\alpha}^\dagger c_{i\alpha} \rangle $ for an antivortex. Combining I and II results in  $\langle  c_{i\alpha}^\dagger c_{i\alpha}\rangle = 1-\langle c_{i\alpha}^\dagger c_{i\alpha} \rangle =1/2.$ The same proof goes for $\langle f_{i\alpha}^\dagger f_{i\alpha}\rangle$.

In the approach we have taken, zero modes play a crucial role -- in their absence, we would have to occupy the negative energy states with up and down spins and end up with a Dirac sea spin {\it singlet} state. To find the zero modes we expand the Hamiltonian near the Dirac nodes, and set energy=0 to obtain  (near the $+\bm{k}_D$ node):
\begin{equation}\label{eq:zero-mode}
\begin{bmatrix}
i\tau_2\partial_1  +  i \tau_1 \partial_2  & b \\
 b^*  & -   \tau_2 (i\partial_1+a_1)  -  \tau_1(i\partial_2+a_2)   \\		
\end{bmatrix}
\pi_+=0,
\end{equation}
where $\pi_+$ is the column vector 
$\begin{pmatrix}
c_{A+}&
c_{B+}&
f_{A+}&
f_{B+}&
\end{pmatrix}^T
$, and $\tau_{\mu}$ are Pauli matrices acting on the AB flavors. $a_i$ are the cartesian coordinates of $\bm{a}(\bm{r}) = a_\theta (r) \hat{\bm{\theta}}$. The ratio $\zeta= t/\chi$ has been eliminated from the zero mode equations after scaling $\begin{pmatrix}
c_{A+}\\
c_{B+}\\ \end{pmatrix} \rightarrow  \begin{pmatrix}
c_{A+}\\
c_{B+}\\ \end{pmatrix}/\sqrt{\zeta}, \  b \rightarrow \sqrt{\zeta} b$. The zero mode equations are studied in the polar coordinates, which is natural in our gauge choice. A further rescaling
\<
\pi_+  \rightarrow \exp\l(-\int_0 ^r  a_{\theta}(\rho) d\rho/2\r) \pi_+,
\>
transforms this problem to

\begin{equation}
\begin{split}
e^{i\theta}\l(\frac{\partial}{\partial r}+\frac{i}{r}\frac{\partial}{\partial \theta} + \frac{a(r)}{2}\r) c_{B+}(\bm{r}) + b(\bm{r}) f_{A+}(\bm{r}) &=0,\\
 b^*(\bm{r}) c_{B+}(\bm{r}) +  e^{-i\theta} \l(\frac{\partial}{\partial r}-\frac{i}{r}\frac{\partial}{\partial \theta} -\frac{a(r)}{2}\r) f_{A+}(\bm{r})&= 0,
\end{split}
\end{equation}
which has been studied recently~\cite{Jackiw08}, and the {\it existence} of 1 decaying zero mode is shown analytically. The same argument goes for the zero mode equations near the $-\bm{k}_D$ mode. Therefore counting $\uparrow$ and $\downarrow$ spin flavors there are 4 zero modes in the KV background. We have also modeled zero mode differential equations numerically  and found  a decaying zero mode in confirmation of the analytical arguments. Furthermore we model the original {\it lattice} model in a KV background, and in diagonalizing the lattice Hamiltonian we find the evidence of very small energy modes in the Kondo gap which rapidly converge to zero as the system size is increased. We have also confirmed that a KV with $2\pi n$ flux has $4n$ zero modes.

\begin{figure}
\subfigure[] 
{
    \label{fig:sub:a}
    \includegraphics[width=4cm]{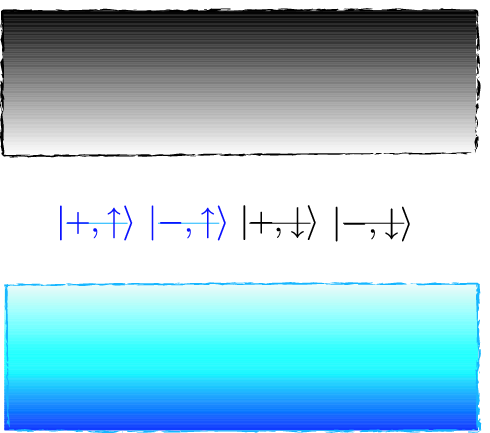}
}
\hspace{0.1cm}
\subfigure[] 
{
    \label{fig:sub:b}
    \includegraphics[width=3cm]{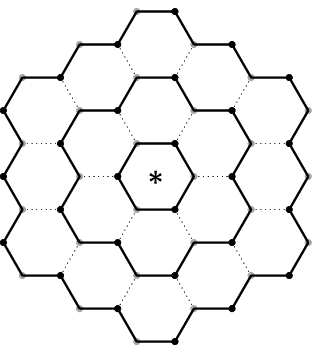}
}
\caption{ \ref{fig:sub:a} The schematic figure of the energy levels of the Hamiltonian of Eq.~(\ref{H_2}) in the presence of a KV background. An example of KV fields (in the continuum limit)  is given in Eq.~(\ref{KVB1}), and Eq.~(\ref{KVB2}). The occupied states are colored by blue. The Dirac sea in the presence of KV has 2 states less than the KI Dirac sea and 2 of the 4 zero modes have to be occupied.
\ref{fig:sub:b} An example of a gauge-symmetric (ring) geometry to study how $v^{i\dagger}_{\xi}$ transforms under the $\pi/3$ rotation around the center of the plaquette labeled $*$. The bold links are the links where the gauge field is non-zero. We let the $2\pi$ flux to spread over a number of rings $N_{\Lambda}$, and choose a symmetric gauge to enclose the flux. The numerics is done in the {\it open} boundary condition.}
\label{fig:sub} 
\end{figure}

After establishing the zero modes, we can now discuss the construction of a spin-1 vortex creation operator.  We define vortex creation operators as operators that {\it increase} the gauge flux  by $2\pi$. If we only limit ourselves with states that are connected to MF state, the $2\pi$ flux-increasing operator contains two terms:
\begin{align}
m_{(a\alpha)(b\beta)}^{+ \dagger} &= z_{a\alpha}^{+\dagger}z_{b\beta}^{+\dagger}\ket{DS,+}\bra{G} \\
m_{(a\alpha)(b\beta)}^{- \dagger} &= \ket{G}\bra{DS,-}z_{b\beta}^{-}z_{a\alpha}^{-},
\end{align}
where $a,b$ $\in$ $\{+,-\}$ are nodal, and $\alpha,\beta \in \{\uparrow,\downarrow\}$ are spin flavors. $\ket{G}$ is the MF ground state, $\ket{DS+}$ is the Dirac sea of {\it negative} energy states in the presence of the $+2\pi$ gauge flux, and $z^{-}_{a\alpha}$ is the zero mode annihilation operator (with $a\alpha$ flavor) for the state with $-2\pi$ gauge flux, etc. The Dirac sea in the presence of a KV has 2 states less than the KI Dirac sea and therefore 2 of the 4 zero modes have to be occupied. There are $\frac{4\times3}{2}$ ways to occupy the zero modes. We classify these 6 KV creation operators into spin-triplet nodal-singlet, and spin-singlet nodal-triplet operators. Since we are interested in the magnetically ordered phases that can arise by condensing Kondo vortices we focus on the spin 1 and nodal singlet operator. The 3 components of this spin triplet vortex creation operator is then given by:
\<
\begin{split}
v^{i\dagger}_{\xi} &= \l[(i\sigma^2)\sigma^i\r]_{\alpha \beta} (i\mu^2)_{ab}  m_{(a\alpha)(b\beta)}^{+\ \dagger} \\&+ \xi \l[\sigma^i (i\sigma^2)\r]_{\alpha \beta} (i\mu^2)_{ab}  m_{(a\alpha)(b\beta)}^{-\ \dagger}
\end{split}
\>
where $\xi$ is an arbitrary constant at this point. It is straightforward to show that under $SU(2)$ rotation in the spin space, $v_\xi^{i\dagger}$ is rotated as a $O(3)$ vector.
\begin{table}[!h]
\begin{tabular}{|c||c|c|c|c|}\hline
\ \ \  & $c_{i\alpha}$ & $f_{i\alpha}$ & $b_i$ & $a_{ij}$ \\
 \hline
$\mathcal{T}$ & $(i\sigma^2)_{\alpha\beta}c_{i\beta}$ & $(i\sigma^2)_{\alpha\beta}f_{i\beta}$ & $b_i^*$ & $-a_{ij}$ \\
\hline
$\mathcal{C}$ & $\epsilon_i (i\sigma^2)_{\alpha\beta}c_{i\beta}^\dagger$ & $-\epsilon_i (i\sigma^2)_{\alpha\beta}f_{i\beta}^\dagger$ & $b_i^*$ & $-a_{ij}$ \\
\hline
$\mathcal{R}^*_{\pi/3}$ & $c_{i'\alpha}$ & $f_{i'\alpha}$ & $b_{i'}$ & $a_{i'j'}$ \\
\hline
\end{tabular}
\caption{The table of the transformation of lattice fields under time-reversal $\mathcal{T}$, charge-conjucation $\mathcal{C}$, and a $\pi/3$ rotation around the center of a plaquette (labeled $*$) $\mathcal{R}^*_{\pi/3}$ . Primed $i'$ etc, is just the transformed index under $\mathcal{R}^*_{\pi/3}$. Other lattice space-group transformations (translations, rotations, and reflections) acts in the same way as in $\mathcal{R}^*_{\pi/3}$ -- in that they only act on the site indices $i\rightarrow i'$.}
\label{table:lattice}
\end{table}

We next wish to see if condensation of the spin triplet $v^{i\dagger}_{\xi}$ can lead to antiferromagnetic order. This  requires that the $v^{i\dagger}_{\xi}$ transform identically to the Neel order parameter under {\it all} symmetries (spin, lattice, time reversal and charge conjugation). The transformation properties of vortex operators under lattice and other symmetries has been addressed  recently in the more complicated cases of {\it gapless} spin liquids. One approach has been to measure spin operators expectation values in the vortex background {\it after projection}~\cite{Hermele08}. Alternatively the transformation of each single particle state in the presence of a vortex background is studied~\cite{Alicea05-06}. The transformation of the vortex creation operators can then be read since vortex creation operators (in this framework) are constructed from the single particle states. In contrast to  the projection approach which is a gauge invariant method, the second approach only works in gauge-symmetric configurations. We adopt the ``single-particle" approach of Alicea {\it et al.} in our problem. In our model we further have the advantage that we study the quantum number of these vortex excitations in a {\it gapped} (KI) phase.

We first focus on the transformation of $v^{i\dagger}_{\xi}$ under time-reversal transformation. Since time-reversal changes $+2\pi$ flux to $-2\pi$ flux it should send $v^{i\dagger}_{\xi}$ to $v^{i}_{\xi}$ {\it except} for possible phase factors.  Demanding this will require $\xi=\pm1$. We sketch the proof here. First we obtain how the individual terms in the definition of $v^{i\dagger}_{\xi}$ transform under time-reversal:
\begin{equation}
\begin{split}
&\mathcal{T}: [(i\sigma^2) \sigma^i]_{\alpha \beta} (i\mu^2)_{ab} m_{(a\alpha)(b\beta)}^{+ \dagger} \\
&\rightarrow [(i\sigma^2) \sigma^{i T}]_{\alpha \beta} (i\mu^2)_{ab} (i\sigma^2)_{\alpha \alpha'} (i\sigma^2)_{\beta\beta'} m_{(a\alpha')(b\beta')}^{-} \\
&= \l[ -(i\sigma^2) (i\sigma^2) \sigma^{i T}  (i\sigma^2) \r]_{\alpha' \beta'}  (i\mu^2)_{ab} m_{(a\alpha')(b\beta')}^{-}\\
&= -[(i\sigma^2) \sigma^i]_{\alpha \beta} (i\mu^2)_{ab} m_{(a\alpha)(b\beta)}^{-}
\end{split}
\end{equation}
Similar algebra for the second term in the definition of $v^{i\dagger}_{\xi}$ results in
\<
\begin{split}
\mathcal{T}: v^{i\dagger}_{\xi} \rightarrow &-[(i\sigma^2) \sigma^i]_{\alpha \beta} (i\mu^2)_{ab} m_{(a\alpha)(b\beta)}^{-}  \\&- \xi^* \l[\sigma^i (i\sigma^2)\r]_{\alpha \beta} (i\mu^2)_{ab}  m_{(a\alpha)(b\beta)}^{+}
\end{split}
\>
Comparing this to $v^{i}_{\xi}$:
\<
\begin{split}
v^{i}_{\xi} &= -\l[\sigma^i (i\sigma^2)\r]_{\alpha \beta} (i\mu^2)_{ab}  m_{(a\alpha)(b\beta)}^{+}\\
 &-\xi^* [(i\sigma^2) \sigma^i]_{\alpha \beta} (i\mu^2)_{ab} m_{(a\alpha)(b\beta)}^{-},
\end{split}
\>
results in $\xi^{*2}=1$, therefore $\xi=\pm1$. Due to anti-unitary character of the time reversal transformation $\pm$ factor can {\it not} be absorbed by a U(1) rotation. Therefore we have two classes of vortex creation operators and their condensation lead to very distinct phases. For example condensation of $v^{i\dagger}_{+}$ can not describe an AF  phase as it is even under time-reversal.  The charge conjugation acts similarly to time reversal in that it sends a vortex to an antivortex. A careful treatment of charge conjugation results in
$\mathcal{C}: \bm{v}_{\pm}^{\dagger}\rightarrow \mp \bm{v}_{\pm}$.

To obtain the transformation of $v^{i\dagger}_{\xi}$ under $\mathcal{R}^*_{\pi/3}$ we resort to numerics. We consider a gauge symmetric configuration [shown in Fig.~\ref{fig:sub:b}] and apply the method of Alicea {\it et al.}~\cite{Alicea05-06}. We find $\mathcal{R}^*_{\pi/3}: v^{i\dagger}_{\xi} \rightarrow -v^{i\dagger}_{\xi}$. The minus sign is obtained independent of lattice sizes and vortex configurations. This is quite a nontrivial result as all the states in the Dirac sea and 2 zero modes contribute to this {\it minus} sign.

\begin{table}[!h]
\begin{tabular}{|l||c|c|c|c|}\hline
 & $\bm{v}_{\pm}^{\dagger}$ & $\bm{v}_{\pm}+\bm{v}_{\pm}^{\dagger}$ &  $i(\bm{v}_{\pm}-\bm{v}_{\pm}^{\dagger})$ & $i\bm{v}_{\pm}\times\bm{v}_{\pm}^{\dagger}$\\
\hline
$\mathcal{T}$  &   $ \pm \bm{v}_{\pm}$ & $\pm$ & $\pm$ & $+$ \\
\hline
$\mathcal{C}$  &   $ \mp \bm{v}_{\pm}$ & $\mp$ & $\pm$ & $-$ \\
\hline
$\mathcal{R}^*_{\pi/3}$ & $ - \bm{v}_{\pm}^{\dagger}$ & $-$ & $-$ & $+$\\
\hline
\end{tabular}
\caption{The table of the transformation of the $O(3)$ vortex creation operator $\bm{v}_{\pm}^{\dagger}$.}
\label{T:KV}
\end{table}

To find how $v^{i\dagger}_{\xi}$ transforms under lattice translations we use  the following identity in the Honeycomb lattice
$R_{\pi/3}^* T_{\bm{a}_1} T_{\bm{a}_2} R_{\pi/3}^{*-1} T_{\bm{a}_2}^{-1} = 1,$
where $\bm{a}_1=(0,\sqrt{3})$, and $\bm{a}_2=(-3/2,-\sqrt{3}/2)$ in the units of nearest neighbor links. This results in $T_{\bm{a}_1}:v^{i\dagger}_{\xi} \rightarrow v^{i\dagger}_{\xi}$. The same result holds for $T_{\bm{a}_2}$. 

Results of the symmetry transformations are summarized in table II. We see that  ${\rm Re}(\bm{v}_{-}) = \bm{v}_{-}+\bm{v}_{-}^\dagger$ transforms identically to the standard two sublattice antiferromagnetic Neel order parameter. Thus its condensation will lead to the usual Neel order.

To describe the {\it universality} of the resulting magnetic phase transition it is convenient to pass to a dual description~\cite{fisher-lee-89,nagaosa-lee-00}  directly in terms of the Kondo vortices. As the Kondo hybridization field $b$ is coupled to a gauge field, its vortices do not have any long range interactions. The dual free energy may then be readily written down by demanding invariance under all physical symmetries and is given by:
\<
\begin{split}F = \sum_{\xi=\pm1} \Bigl(t_\xi |\bm{v}_{\xi}|^2+r_\xi(\bm{v}_{\xi}^2+\bm{v}_{\xi}^{*2})+u_\xi |\bm{v}_{\xi}|^4\\
+s_\xi \bm{v}_{\xi}^2 \bm{v}_{\xi}^{*2}+ w_\xi |\bm{v}_{\xi}\times \bm{v}_{\xi}^*|^2+\cdots \Bigr).
\end{split}
\>
We emphasize that, in contrast to the usual boson-vortex duality, due to the $r_\xi$ terms here the vorticity is not conserved. In other words the free energy is {\it not} invariant under a phase rotation of the vortex fields. This is because the gauge field $a_{ij}$ in the original description is compact. This allows for instanton configurations where the gauge flux can change in units of $2\pi$. However the spin carried by the vortices prohibits single instanton events; pairs of vortices in a spin singlet can nevertheless be created or destroyed as described by the $r_\xi$ term. 
 
If $r_-<0$,  ${\rm Re}(\bm{v}_{-})$ will condense first, while ${\rm Im}(\bm{v}_{-})$ remains zero and the transition is then described by the following free energy
\<
\begin{split}F = (t_-+r_-){\rm Re}(\bm{v}_{-})^2 + (u_-+s_-) {\rm Re}(\bm{v}_{-})^4+\cdots,
\end{split}
\>
which describes an $O(3)$ transition for ${\rm Re}(\bm{v}_{-})$. Therefore in our theoretical framework, this KV mediated AF transition is an $O(3)$ transition. This is perhaps not unexpected, because a charge gap exists on both sides of the AF transition and the notion of an onset of Kondo screening is an artifact of mean field theory.

\begin{table}[!h]
\begin{tabular}{|l||c|c|c|}\hline
 & $\Pi$  & $\bar{\Pi} \sigma^i \kappa^3 \mu^3 \Pi$ & $\bar{\Pi} \sigma^i \kappa^3 \Pi$ \\
\hline
{$\mathcal{T}$} & $-\sigma^2 \mu^2 \tau^2 \kappa^3 \Pi$ & $-$ & $+$ \\
\hline
{$\mathcal{C}$} & $-i(\bar{\Pi}\sigma^2\mu^1\tau^2)^T$ & $+$ & $-$ \\
\hline
$\mathcal{R}^*_{\pi/3}$ & $-\tau^3 \exp(-i\frac{\pi}{3} \tau^3)\exp(+i\frac{\pi}{3} \mu^3) \mu^2   \Pi$ & $-$ & $+$\\
\hline
\end{tabular}
\caption{ The transformation of the 8 component fermionic field $\Pi$ near the Dirac nodes. $\bar{\Pi}= \Pi^\dagger \tau^3$. $\kappa^i$, $\sigma^i$, $\mu^i$, and $\tau^i$ are Pauli matrices acting on  fermionic $\Psi\Phi$, spin $\uparrow\downarrow$, nodal $+-$, and sublattice $AB$ flavors respectively (see Table \ref{field-def}).}
\label{low-energy-fields}
\end{table}

\begin{table}[!h]
\begin{tabular}{|c|c|c|c|c|c|}
\hline
 $\Pi$ &  $\Psi$ & $\Psi_{\uparrow}$ & $\psi_{+\uparrow}$ & $\Phi_{\uparrow}$ & $\phi_{+ \uparrow}$  \\
\hline
$\begin{pmatrix} \Psi \\ \Phi \end{pmatrix}$ &
$\begin{pmatrix} \Psi_{\uparrow} \\ \Psi_\downarrow \end{pmatrix}$ &
$\begin{pmatrix} \psi_{+ \uparrow} \\ \tau^2 \psi_{-\uparrow} \end{pmatrix}$&
$\begin{pmatrix} c_{A+ \uparrow} \\  c_{B+ \uparrow}  \end{pmatrix}$&
$\begin{pmatrix} \tau^3 \phi_{+ \uparrow} \\ -i \tau^1 \phi_{-\uparrow} \end{pmatrix}$&
$\begin{pmatrix} f_{A+ \uparrow} \\  f_{B+ \uparrow }  \end{pmatrix}$\\
\hline
\end{tabular}
\caption{ The definition  of the 8 component fermionic field $\Pi$. Redundant definitions are ignored.}
\label{field-def}
\end{table}

The phase characterized  by $i \bm{v}_{\xi}\times \bm{v}_{\xi}^* = {\rm Re}(\bm{v}_{\xi}) \times  {\rm Im}(\bm{v}_{\xi}) \neq 0$ is also a very interesting phase. This is the analogue of the Kane-Mele~\cite{kane-mele} phase in our model, since $i \bm{v}_{\xi}\times \bm{v}_{\xi}^*$ transforms the same way as $\bar{\Pi} \sigma^i \kappa^3 \Pi$ (see Tables~\ref{T:KV}, \ref{low-energy-fields}, and \ref{field-def}), and the gap generated by $\bar{\Pi} \sigma^i \kappa^3 \Pi$ supports a spin Hall effect.  This can be seen by applying the external gauge fields $A_\mu^c$ and $
A_\mu^s$ which are coupled to charge and spin currents. The effective action for the gauge fields is obtained after integrating out the fermions $S_{eff}[A_\mu^c,A_\mu^s]=\frac{i}{2\pi}\int d^3x \epsilon_{\mu\nu\lambda}A_\mu^c \partial_\nu A_\lambda^s$ indicating the spin Hall conductivity of magnitude $\frac{1}{2\pi}$~($e=1$) \cite{kane-mele,tarun08}. This phase is charactrerized by the spontaneous onset of spin-orbit coupling~\cite{kane-mele}. However to condense ${\rm Re}(\bm{v}_{\xi}) \times  {\rm Im}(\bm{v}_{\xi})$  we should tune {\it both} $t_\xi$ and  $r_\xi$, and the transition from KI to this Kane-Mele phase is controlled by a multi-critical point.

The KV excitation we considered is a charge-0 spin-1 excitation and deep in the KI phase is not distinguishable from a particle-hole excitation which is accompanied by a spin-flip. However we have checked the energetics of the KV excitation and have found that the KV excitation costs less energy than a simple KI particle-hole excitation. Therefore in the band theory terminology (as far as long distance effects of these excitations are concerned) the KV (in our model) may correspond to triplet {\it excitonic} bound states in the KI gap and the onset of AF may be viewed as the condensation of these triplet excitons.

In summary we go beyond mean field theory to describe the destruction of the Kondo admixture parameter due to the proliferation of Kondo vertices. We find that the vortex field carries spin 1 quantum numbers and its condensation leads to AF ordering. In this way the suppression of Kondo screening is linked to the appearance of AF. The drawback of our model is that a charge gap always exist, both in the KI and in the AF phase and is nonzero at the onset of AF. As a result there is no sharp onset for the KI, in contrast to the HFL case where the Knodo phase is characterized by the sharp onset of a large Fermi surface. Nevertheless, we hope that this line of investigation might shed some light on the more difficult problem of the AF-HFL quantum critical point. 

{\it Acknowledgments}---We acknowledge helpful conversations with Jason Alicea. SS is specially grateful for the conversations he had with Ying Ran, Michael Hermele, Ashvin Vishwanath, and Roman Jackiw.  PAL was supported by the Department of Energy under grant DE-FG02-03ER46076, and TS by NSF Grant DMR-0705255.

\end{document}